\documentstyle[12pt,epsf]{article} \hoffset -0.8in \textwidth 7.00in
\textheight
8.5in
\newskip\humongous \humongous=0pt plus 1000pt minus 1000pt
\def\caja{\mathsurround=0pt}
\def\eqalign#1{\,\vcenter{\openup1\jot \caja
        \ialign{\strut \hfil$\displaystyle{##}$&$
        \displaystyle{{}##}$\hfil\crcr#1\crcr}}\,}
\newif\ifdtup

\def\eqright #1\cr{\noalign{\hfill$\displaystyle{{}#1}$}}
\def\eqleft #1\cr{\noalign{\noindent$\displaystyle{{}#1}$\hfill}}

\def\oldreffmt#1{\rlap{[#1]} \hbox to 2\parindent{}}

\def\figfmt#1{\rlap{Figure {#1}} \hbox to 1in{}}

\def\auto{\eqno(\refstepcounter{equation}\theequation)}
\def\begineq #1\endeq{$$ \refstepcounter{equation}\eqalign{#1}\eqno
	(\theequation) $$}
\def\contlimit{\,{\hbox{$\longrightarrow$}\kern-1.8em\lower1ex
\hbox{${\scriptstyle (a\rightarrow0)}$}}\,}
\def\centeron#1#2{{\setbox0=\hbox{#1}\setbox1=\hbox{#2}\ifdim
\wd1>\wd0\kern.5\wd1\kern-.5\wd0\fi
\copy0\kern-.5\wd0\kern-.5\wd1\copy1\ifdim\wd0>\wd1
\kern.5\wd0\kern-.5\wd1\fi}}
\def\centerover#1#2{\centeron{#1}{\setbox0=\hbox{#1}\setbox
1=\hbox{#2}\raise\ht0\hbox{\raise\dp1\hbox{\copy1}}}}
\def\centerunder#1#2{\centeron{#1}{\setbox0=\hbox{#1}\setbox
1=\hbox{#2}\lower\dp0\hbox{\lower\ht1\hbox{\copy1}}}}
\def\lsim{\;\centeron{\raise.35ex\hbox{$<$}}{\lower.65ex\hbox
{$\sim$}}\;}
\def\gsim{\;\centeron{\raise.35ex\hbox{$>$}}{\lower.65ex\hbox
{$\sim$}}\;}

\def\super#1{\ifmmode \hbox{\textsuper{#1}}\else\textsuper{#1}\fi}
\def\textsuper#1{\newcount\holdspacefactor\holdspacefactor=\spacefactor
$^{#1}$\spacefactor=\holdspacefactor}

\def\getcite#1,{\advance\citenumber by1
\def\getcitearg{#1}\def\lastarg{@}
\ifnum\citenumber=1
\ref{#1}\let\next=\getcite\else\ifx\getcitearg\lastarg\let\next=\relax
\else ,\ref{#1}\let\next=\getcite\fi\fi\next}

\def\pom{{\rm P\kern -0.53em\llap I\,}}
\def\spom{{\rm P\kern -0.36em\llap \small I\,}}
\def\sspom{{\rm P\kern -0.33em\llap \footnotesize I\,}}

\relax

\newskip\humongous \humongous=0pt plus 1000pt minus 1000pt
\def\caja{\mathsurround=0pt}
\def\eqalign#1{\,\vcenter{\openup1\jot \caja
        \ialign{\strut \hfil$\displaystyle{##}$&$
        \displaystyle{{}##}$\hfil\crcr#1\crcr}}\,}
\newif\ifdtup

\def\eqright #1\cr{\noalign{\hfill$\displaystyle{{}#1}$}}
\def\eqleft #1\cr{\noalign{\noindent$\displaystyle{{}#1}$\hfill}}

\def\oldreffmt#1{\rlap{[#1]} \hbox to 2\parindent{}}

\def\figfmt#1{\rlap{Figure {#1}} \hbox to 1in{}}

\def\auto{\eqno(\refstepcounter{equation}\theequation)}
\def\begineq #1\endeq{$$ \refstepcounter{equation}\eqalign{#1}\eqno
	(\theequation) $$}
\def\contlimit{\,{\hbox{$\longrightarrow$}\kern-1.8em\lower1ex
\hbox{${\scriptstyle (a\rightarrow0)}$}}\,}
\def\centeron#1#2{{\setbox0=\hbox{#1}\setbox1=\hbox{#2}\ifdim
\wd1>\wd0\kern.5\wd1\kern-.5\wd0\fi
\copy0\kern-.5\wd0\kern-.5\wd1\copy1\ifdim\wd0>\wd1
\kern.5\wd0\kern-.5\wd1\fi}}
\def\centerover#1#2{\centeron{#1}{\setbox0=\hbox{#1}\setbox
1=\hbox{#2}\raise\ht0\hbox{\raise\dp1\hbox{\copy1}}}}
\def\centerunder#1#2{\centeron{#1}{\setbox0=\hbox{#1}\setbox
1=\hbox{#2}\lower\dp0\hbox{\lower\ht1\hbox{\copy1}}}}
\def\lsim{\;\centeron{\raise.35ex\hbox{$<$}}{\lower.65ex\hbox
{$\sim$}}\;}
\def\gsim{\;\centeron{\raise.35ex\hbox{$>$}}{\lower.65ex\hbox
{$\sim$}}\;}

\def\super#1{\ifmmode \hbox{\textsuper{#1}}\else\textsuper{#1}\fi}
\def\textsuper#1{\newcount\holdspacefactor\holdspacefactor=\spacefactor
$^{#1}$\spacefactor=\holdspacefactor}

\def\getcite#1,{\advance\citenumber by1
\ifnum\citenumber=1
\ref{#1}\let\next=\getcite\else\ifx#1@\let\next=\relax
\else ,\ref{#1}\let\next=\getcite\fi\fi\next}

\def\upon #1/#2 {{\textstyle{#1\over #2}}}
\relax

\def\til#1{\centeron{\hbox{$#1$}}{\lower 2ex\hbox{$\char'176$}}}
\def\tild#1{\centeron{\hbox{$\,#1$}}{\lower 2.5ex\hbox{$\char'176$}}}
\def\sumtil{\centeron{\hbox{$\displaystyle\sum$}}{\lower
-1.5ex\hbox{$\widetilde{\phantom{xx}}$}}}

\def\pom{{\rm P\kern -0.53em\llap I\,}}
\def\spom{{\rm P\kern -0.36em\llap \small I\,}}
\def\sspom{{\rm P\kern -0.33em\llap \footnotesize I\,}}

\newcommand{\bit}{\begin{itemize}}
\newcommand{\eit}{\end{itemize}}

\newcommand{\beq}{\begin{equation}}
\newcommand{\eeq}{\end{equation}}
\newcommand{\beqa}{\begin{eqnarray}}
\newcommand{\eeqa}{\end{eqnarray}}

\begin{document}

\begin{titlepage}
\rightline{\vbox{\halign{&#\hfil\cr
&ANL-HEP-CP-95-??\cr }}}

{}~
\vspace{1in}

\begin{center}

{\bf CONFORMALLY SYMMETRIC CONTRIBUTIONS TO BFKL EVOLUTION AT
NEXT TO LEADING ORDER}
\medskip

Claudio Corian\`{o} and Alan R. White
\footnote{Work supported by the U.S. Department of Energy, Division of High
Energy Physics, Contract\newline W-31-109-ENG-38}
\footnote{Presented by C. Corian\`{o}
at the XXXth Rencontres de Moriond, Les Arcs, France,
March 19-26, 1995.}
\\
\smallskip
High Energy Physics Division, Argonne National Laboratory,
Argonne, IL 60439.

\end{center}

\begin{abstract}

Unitarity corrections to the BFKL evolution at next to leading order
determine a new component of the evolution kernel which is shown to
possess conformal invariance properties. Expressions for the complete spectrum
of the new component and the correction to the intercept of the pomeron
trajectory are presented.
\end{abstract}

\end{titlepage}

The program of resummation of the leading $(\sim \alpha_s Log\,\,1/x)^n$
 and next-to-leading order
corrections to the small-x behaviour of the gluon structure functions
has undergone a revival of interest, mainly motivated by the new experimental
results on DIS at HERA, which seem to be related to
pomeron exchange. The small-x region of QCD, however, merges with the
{\em softer} ``Regge limit of QCD" at small transverse momenta, and it
is this second limit which
can be better described by reggeon unitarity in the $t$-channel\cite{cw1}.
Summation of the next to leading
logs $(\sim [c_1 \alpha_s + c_2 \alpha_s^2]~ Log\,\,1/x)^n$ in the
Regge limit has special features connected to the phenomenon of reggeization
of the gluon. Reggeization can be proven in perturbation theory and its
effects can be ``iterated" in the construction of an evolution kernel using
perturbative s-channel unitarity, as done by Lipatov and collaborators
and by Bartels\cite{cw1}. Our results are derived by imposing conditions
of $t$-channel unitarity on the scattering amplitude in the 2-reggeon sector.
A new scale invariant component of the BFKL evolution is identified, which is
expressed in terms of massless transverse momentum diagrams \cite{cw1}.
The new component is given by
$$
\eqalign{{1 \over (g^2N)^2} K^{(4n)} ~=~ \Biggl( ~{\cal K}_0~\Biggr)
{}~+~\Biggl( ~{\cal K}_1~\Biggr) ~-~\Biggl( ~{\cal K}_2~\Biggr)
, }
\auto\label{4n}
$$
where
$$
\eqalign{ {\cal K}_0~=~{1 \over 8\pi}{\eta}^2(k^2)^D \biggl(\delta^2(k-k')
+\delta^2(k-k')\biggr)}
\auto
$$
$$
\eqalign{ {\cal K}_1~=~&{\eta \over 8\pi^2} \biggl( 2 k^2
{k'}^2([(k'-k)^2]^{D/2-2}~+~[(k+k')^2]^{D/2-2})~\cr
&-(k^2 [{k'}^2]^{D/2} ~+~[k^2]^{D/2} {k'}^2)\bigl({1 \over (k'-k)^2}
{}~+~ {1 \over(k'+k)^2}\bigr)\biggr)}
\auto
$$
and
\beqa
{\cal K}_2~=~{\eta \over 4\pi^2} ~ {k^2 {k'}^2 (k^2-{k'}^2)\over
(k+k')^2 (k-k')^2}\left( (k^2)^{D/2~-1} -
({k'}^2)^{D/2~-1}\right).
\eeqa
We have set $D=2 +\epsilon$ and $\eta=2/(D-2)$. $\Biggl( ~{\cal K}_0~\Biggr)
{}~+~\Biggl( ~{\cal K}_1~\Biggr)$ can be rexpressed in terms of the square of
the lowest order Lipatov kernel. $\Biggl( ~{\cal K}_2~\Biggr) $ is a new
component which is separately infrared safe and has distinct conformal
invariance properties.

The complete eigenvalue spectrum of $K^{(4n)}$ is given
by $ N^2g^4 {\cal E}(\nu,n)$ where
$$
{\cal E}(\nu,n)~=~{1 \over \pi} [\chi(\nu,n)]^2~-~\Lambda(\nu,n) ~.
\auto
$$
$\chi(\nu,n)$ is the Lipatov characteristic function giving the $O(g^2)$
eigenvalues. $\Lambda(\nu,n)$ is due to the new ${\cal K}_2$ component and
is given by
$$
\eqalign{ \Lambda(\nu,n) ~=~-~{1 \over 4\pi}
\biggl(\beta'\bigl({|n| + 1\over 2} +
i\nu\bigr)
{}~+~\beta'\bigl({|n| + 1 \over 2} -i\nu\bigr)\biggr). }
\auto\label{lam}
$$
where
$$
\eqalign{\beta'(x)~=~{1\over 4}\biggl(\psi'\bigl({x+1\over 2}\bigr) -
\psi'\bigl({x\over 2}\bigr)\biggr) , }
\auto\label{ps1}
$$
with
$$
\eqalign { \psi'(z)~=~\sum_{r=0}^{\infty} {1
\over (r+z)^2}, }
\auto\label{ps2}
$$
from which it follows that the eigenvalues $\Lambda(\nu,n)$ are all real.
We can also write \cite{cw1}
$$
\eqalign{\Lambda(\nu,n) ~&=~-~{1 \over 8\pi}\biggl(\beta'\bigl(m \bigr) ~+~
\beta'\bigl(1-m \bigr)~+~\beta'\bigl(1-\tilde{m} \bigr)
{}~+~\beta'\bigl(\tilde{m} \bigr)\biggr) \cr
&\equiv ~{\cal G}\bigl[m(1-m)\bigr]~+
{}~{\cal G}\bigl[\tilde{m}(1-\tilde{m})\bigr] }
\auto
$$
where $m~=~1/2+i\nu +n/2$ and  $\tilde{m}~=~1/2+i\nu -n/2$.
This is the holomorphic factorization property satisfied by the leading
order eigenvalues $\chi(\nu,n)$, which is directly related to conformal
invariance.

Finally we discuss the numerical values that we obtain from our results.
The leading eigenvalue is at $\nu=n=0$, as it is for the $O(g^2)$ kernel.
{}From ${\cal K}_2$ alone, the correction to the Pomeron intercept $\alpha_0$
is (we recall that $\alpha_0 - 1$ gives the inverse power behavior of $F_2(x)$)
$$
\eqalign{ {9g^4\over 16\pi^3}\Lambda(0,0)
{}~\sim~-16.3 {{\alpha_s}^2 \over \pi^2} }
\auto
$$
The complete $\hat{K}^{(4n)}$ gives
$$
\eqalign{{\cal E}(0,0)/(16\pi^3)
{}~\sim&~{N^2g^4 \over 16 {\pi}^4}\biggl([2ln2]^2
{}~-~1.81 \biggr)\cr
{}~\sim&~{9g^4 \over 16{\pi}^4}\times 0.11 ~\sim~ {{\alpha_s}^2 \over
\pi^2} }
\auto
$$
giving a very small positive effect. However, the disconnected parts of
$K^{(4n)}$ do not have a consistent reggeization interpretation\cite{cw1}.
To obtain a consistent scale-invariant $O(g^4)$ kernel it is necessary to
subtract the square of the leading-order kernel. This gives, as a
modification of $\alpha_0$,
$$
\eqalign{{\tilde{{\cal E}}(0,0) \over 16\pi^3}
{}~=&~{N^2g^4 \over 16 {\pi}^4}\biggl(
-3 [\chi(0,0)]^2
{}~-~\Lambda(0,0) \biggr)\cr
\sim &~{9g^4 \over 16{\pi}^4}\times  (-5.76  - 1.81)\cr
\sim&~-68 {{\alpha_s}^2 \over \pi^2}}
\auto
$$
which is a substantial negative correction. The scale-invariant kernel should
accurately describe the full next-to-leading-order kernel in the transverse
momentum infra-red region. We conclude that this region can produce a strong
reduction of the BFKL small-x behavior.

\vspace{1cm}
\centerline{\bf Acknowledgements}
C.C. thanks A. Capella, C. I. Tan and G. Korchemsky for discussions.


\end{document}